\documentclass[10pt,english]{article}
\usepackage{times}
\usepackage[T1]{fontenc}
\usepackage[latin1]{inputenc}
\usepackage{amsmath}
\usepackage{amssymb}

\makeatletter

\newcommand{\noun}[1]{\textsc{#1}}

\usepackage{slashed}

\usepackage{babel}
\makeatother
\begin{document}

\title{\textbf{A straightforward realization of a quasi-inverse seesaw mechanism
at TeV scale}}

\author{\noun{ADRIAN} PALCU}

\date{\emph{Faculty of Exact Sciences - ''Aurel Vlaicu'' University of
Arad, Str. Elena Dr\u{a}goi 2, 310330 - Arad, Romania}}

\maketitle
\begin{abstract}
In this letter we address the issue of generating appropriate tiny
neutrino masses within the framework of a particular $SU(3)_{c}\otimes SU(3)_{L}\otimes U(1)_{X}$
gauge model by adding three singlet exotic Majorana neutrinos to the
ones included in the three lepton triplet representations. The theoretical
device is the general method of treating gauge models with high symmetries
$SU(3)_{c}\otimes SU(N)_{L}\otimes U(1)_{X}$ proposed by Cot\u{a}escu
more than a decade ago. When it is worked-out in the 3-3-1 model it
supplies a unique free parameter ($a$) to be tuned in order to get
a realistic mass spectrum for both the boson and charged-fermion sectors.
Its most appealing feature (of special interest here) is that it contains
all the needed ingredients to realize the inverse seesaw mechanism
for neutrinos. The mandatory couplings leading to the lepton number
soft violation in pure Majorana terms result without invoking any
element outside the model (such as GUT scales, as one usually does
in the literature). The overall breaking scale in this particular
model can be set around 1 TeV so its phenomenology is quite testable
at present facilities.

PACS numbers: 14.60.St; 12.60.Cn; 12.60.Fr; 14.80.Cp.

Key words: inverse seesaw mechanism, right-handed neutrinos, extensions
of the SM.
\end{abstract}

\section{Introduction}

It is well-known that the Standard Model (SM) (\cite{key-1} - \cite{key-3})
- based on the gauge group $SU(3)_{c}\otimes SU(2)_{L}\otimes U(1)_{Y}$
undergoing a spontaneous symmetry breaking (SSB) in its electro-weak
sector up to the universal $U(1)_{em}$ - is not a sufficient device,
at least for some stringent issues in the particle physics today.
When it comes to generating neutrino tiny masses \cite{key-4,key-5},
the framework of the SM is lacking the needed ingredients, so one
should call for some extra considerations which are less natural in
the context. One of the ways out seems to be the enlargement of the
gauge group of the theory as to include naturally among its fermion
representations some right-handed neutrinos - mandatory.elements for
some plausible mass terms in the neutrino sector Yukawa Lagrangian
density (Ld) of the theory. 

Among such possible extensions of the SM, the so called ''3-3-1''
class of models \cite{key-6} - \cite{key-9} - where the new gauge
group is $SU(3)_{c}\otimes SU(3)_{L}\otimes U(1)_{X}$ - has meanwhile
established itself as a much suitable candidate. A systematic classification
\cite{key-10} - \cite{key-12} and phenomenological study of these
models (especially those which don't include exotic electric charges
\cite{key-13} - \cite{key-23}) )have been done during the last two
decades. Some of the studies address the neutrino mass issue \cite{key-24}
- \cite{key-35} with viable results within the framework of such
models. 

Here we propose a slightly different approach from the canonical one,
in the sense that we apply the prescriptions of the general method
\cite{key-36} of treating gauge models with high symmetries. Proposed
initially by Cot\emph{\u{a}}escu, it essentially consists of a general
algebraical procedure in which electro-weak gauge models with high
symmetries ($SU(N)_{L}\otimes U(1)_{Y}$) achieve their SSB in only
one step up to the residual $U(1)_{em}$ by means of a special Higgs
mechanism. The scalar sector is organized as a complex vector space
where a real scalar field $\varphi$ is introduced as the norm for
the scalar product among scalar multiplets. It also ensures the orthogonality
in the scalar vector space. Thus, the survival of some unwanted Goldstone
bosons is avoided. This leads to a one-parameter mass spectrum, due
to a restricting trace condition that has to hold throughout. The
compatibility of this particular method with the canonical approach
to 3-3-1 models in the literature was proved in a recent paper by
the author\cite{key-37} where an appealing outcome with only two
physical massive Higgses with non-zero interactions finally emerged.
This will be precisely the framework of our proceedings here. Furthermore,
we exploit the realization of a kind of quasi-inverse seesaw mechanism
\cite{key-38} - \cite{key-46} in the framework of 3-3-1 gauge models
with 3 right-handed neutrinos $\left(\nu_{R}\right)$ included in
the fermion triplets and 3 exotic sterile Majorana singlets $\left(N_{R}\right)$
, in which the free parameter (let's call it $a$) is tuned in order
to obtain the whole mass spectrum. An apparently unused up to now
parameter $\eta_{0}$ in the general method proves itself here as
the much needed ''lepton number violating'' coupling to achieve
the Majorana mass terms for $N_{R}$ in the neutrino sector.

The letter is divided into 5 sections. It begins with a brief presentation
of the model and its parametrization supplied by the general Cot\u{a}escu
method (in Sec.2) and continues with the inverse seesaw mechanism
worked out within this framework (Sec. 3) and the tuning of the parameters
(Sec. 4) in order to obtain phenomenologically viable results for
the neutrino masses. Some conclusions are sketched in the last section
(Sec. 5).

\section{Brief review of the model}

The particle content of the 3-3-1 gauge model of interest here is
the following:

\textbf{Lepton families}

\begin{equation}
\begin{array}{ccccc}
l_{\alpha L}=\left(\begin{array}{c}
\nu_{\alpha}^{c}\\
\nu_{\alpha}\\
e_{\alpha}\end{array}\right)_{L}\sim(\mathbf{1,3},-1/3) &  &  &  & e_{\alpha R}\sim(\mathbf{1},\mathbf{1},-1)\end{array}\label{Eq.1}\end{equation}

\textbf{Quark families}

\begin{equation}
\begin{array}{ccc}
Q_{iL}=\left(\begin{array}{c}
D_{i}\\
-d_{i}\\
u_{i}\end{array}\right)_{L}\sim(\mathbf{3,3^{*}},0) &  & Q_{3L}=\left(\begin{array}{c}
U_{3}\\
u_{3}\\
d_{3}\end{array}\right)_{L}\sim(\mathbf{3},\mathbf{3},+1/3)\end{array}\label{Eq.2}\end{equation}

\begin{equation}
\begin{array}{ccc}
d_{iR},d_{3R}\sim(\mathbf{3},\mathbf{1},-1/3) &  & u_{iR},u_{3R}\sim(\mathbf{3},\mathbf{1},+2/3)\end{array}\label{Eq.3}\end{equation}

\begin{equation}
\begin{array}{ccccccccc}
U_{3R}\sim(\mathbf{3,1},+2/3) &  &  &  &  &  &  &  & D_{iR}\sim(\mathbf{3,1},-1/3)\end{array}\label{Eq.4}\end{equation}
with $i=1,2$. 

The above representations ensure the cancellation of all the axial
anomalies (by an interplay between families, although each one remains
anomalous by itself). In this way one prevents the model from compromising
its renormalizability by triangle diagrams. The capital letters are
reserved foe the exotic quarks ($D_{i}$, $D_{2}$ and $U_{3}$) in
each family. They are heavier than the ordinary quarks known from
the SM. 

To this fermion content one can add 3 Majorana exotic neutrinos $N_{R}\sim(\mathbf{1},\mathbf{1},0)$
without the danger of spoiling the renormalizability. The advantage
these 3 exotic neutrinos bring is that they can play a crucial role
in realizing the inverse seesaw mechanism\cite{key-38} - \cite{key-46}.

\textbf{Gauge bosons}

The gauge bosons of the model are determined by the generators of
the electro-weak $su(3)$ Lie algebra, expressed by the usual Gell-Mann
matrices $T_{a}=\lambda_{a}/2$ . So, the Hermitian diagonal generators
of the Cartan sub-algebra are \begin{equation}
D_{1}=T_{3}=\frac{1}{2}{\textrm{Diag}}(1,-1,0)\,,\quad D_{2}=T_{8}=\frac{1}{2\sqrt{3}}\,{\textrm{Diag}}(1,1,-2)\,.\label{Eq,5}\end{equation}
 In this basis the gauge fields are expressed by: $A_{\mu}^{0}$ (corresponding
to the Lie algebra of the group $U(1)_{X}$) and $A_{\mu}\in su(3)$,
that can be put as \begin{equation}
A_{\mu}=\frac{1}{2}\left(\begin{array}{ccc}
A_{\mu}^{3}+A_{\mu}^{8}/\sqrt{3} & \sqrt{2}X_{\mu} & \sqrt{2}Y_{\mu}\\
\\\sqrt{2}X_{\mu}^{*} & -A_{\mu}^{3}+A_{\mu}^{8}/\sqrt{3} & \sqrt{2}W_{\mu}\\
\\\sqrt{2}Y_{\mu}^{*} & \sqrt{2}W_{\mu}^{*} & -2A_{\mu}^{8}/\sqrt{3}\end{array}\right),\label{Eq.6}\end{equation}
 where $\sqrt{2}W_{\mu}^{\pm}=A_{\mu}^{6}\mp iA_{\mu}^{7}$, $\sqrt{2}Y_{\mu}^{\pm}=A_{\mu}^{4}\pm iA_{\mu}^{5}$,
and $\sqrt{2}X_{\mu}=A_{\mu}^{1}\pm iA_{\mu}^{2}$, respectively.
One notes that apart from the charged Weinberg bosons ($W$) from
SM, there are two new complex boson fields, $X$ (neutral) and $Y$
(charged) - off-diagonal entries in eq.(\ref{Eq.6}). 

The diagonal Hermitian generators will provide us with the neutral
gauge bosons $A_{\mu}^{em}$, $Z_{\mu}$and $Z_{\mu}^{\prime}$. Therefore,
on the diagonal terms in eq.(\ref{Eq.6}) a generalized Weinberg transformation
(gWt) must be performed in order to consequently separate the massless
electromagnetic field from the other two neutral massive fields. One
of the two massive neutral fields is nothing but the $Z^{0}$-boson
of the SM. The details of the general procedure with gWt can be found
in Ref. \cite{key-36} and its concrete realization in the model of
interest here in Refs. \cite{key-19,key-26}. In Ref. \cite{key-19}
the neutral currents for both $Z_{\mu}$and $Z_{\mu}^{\prime}$ are
completely determined, while in Ref. \cite{key-26} the boson mass
spectrum is calculated.

For the sake of completeness we write down the electric charge operator
in this particular 3-3-1 model when Cot\u{a}escu method is involved.
It stands simply as: $Q^{\rho}=\frac{2}{\sqrt{3}}T_{8}^{\rho}+Y^{\rho}$
for each representation $\rho$.

\textbf{Scalar sector and spontaneous symmetry breaking}

In the general method \cite{key-36}, the scalar sector of any $SU(N)_{L}\otimes U(1)_{Y}$
electro-weak gauge model must consist of $n$ Higgs multiplets $\phi^{(1)}$,
$\phi^{(2)}$, ... , $\phi^{(n)}$ satisfying the orthogonal condition
$\phi^{(i)+}\phi^{(j)}=\varphi^{2}\delta_{ij}$ in order to eliminate
unwanted Goldstone bosons that could survive the SSB. Here $\varphi\sim\left(1,1,0\right)$
is a gauge-invariant real field acting as a norm in the scalar space
and $n$ is the dimension of the fundamental irreducible representation
of the gauge group. The parameter matrix $\eta=\left(\eta_{0},\eta{}_{1},\eta{}_{2}..,\eta{}_{n}\right)$
with the property $Tr\eta^{2}=1-\eta_{0}^{2}$ is a key ingredient
of the method: it is introduced in order to obtain a non-degenerate
boson mass spectrum. Obviously, $\eta_{0},\eta{}_{i}\in[0,1)$. Then,
the Higgs Ld reads:

\begin{equation}
\mathcal{L}_{H}=\frac{1}{2}\eta_{0}^{2}\partial_{\mu}\varphi\partial^{\mu}\varphi+\frac{1}{2}\sum_{i=1}^{n}\eta_{i}^{2}\left(D_{\mu}\phi^{(i)}\right)^{+}\left(D^{\mu}\phi^{(i)}\right)-V(\phi^{(i)})\label{Eq. 7}\end{equation}
where $D_{\mu}\phi^{(i)}=\partial_{\mu}\phi^{(i)}-i(gA_{\mu}+g^{\prime}y^{(i)}A_{\mu}^{0})\phi^{(i)}$
act as covariant derivatives of the model. $g$ and $g^{\prime}$
are the coupling constants of the groups $SU(N)_{L}$ and $U(1)_{X}$
respectively. Real characters $y^{(i)}$ stand as a kind of hyper-charge
of the new theory. 

For the particular 3-3-1 model under consideration here the most general
choice of parameters is given by the matrix $\eta^{2}=\left(1-\eta_{0}^{2}\right)Diag\left[1-a\,,\,\frac{1}{2}\left(a-b\right)\,,\,\frac{1}{2}\left(a+b\right)\right]$.
It obviously meets the trace condition required by the general method
for any $a,b\in[0,1)$. After imposing the phenomenological condition
$M_{Z}^{2}=M_{W}^{2}/\cos^{2}\theta_{W}$ (confirmed at the SM level)
the procedure of diagonalizing the neutral boson mass matrix \cite{key-19,key-26}
reduces to one the number of parameters, so that the parameter matrix
reads $\eta^{2}=\left(1-\eta_{0}^{2}\right)Diag\left[1-a\,,\, a\frac{\left(1-\tan^{2}\theta_{W}\right)}{2}\,,\, a\frac{1}{2\cos^{2}\theta_{W}}\right]$. 

With the following content in the scalar sector of the 3-3-1 model
of interest here (and based on the redefinition of the scalar triplets
from the general method, as in the Ref.\cite{key-37})

\begin{equation}
\rho=\left(\begin{array}{c}
\rho_{1}^{0}\\
\\\rho_{2}^{0}\\
\\\rho_{3}^{-}\end{array}\right),\chi=\left(\begin{array}{c}
\chi_{1}^{0}\\
\\\chi_{2}^{0}\\
\\\chi_{3}^{-}\end{array}\right)\sim(\mathbf{1},\mathbf{3},-1/3)\,,\quad\phi=\left(\begin{array}{c}
\phi_{1}^{+}\\
\\\phi_{2}^{+}\\
\\\phi_{3}^{0}\end{array}\right)\sim(\mathbf{1},\mathbf{3},+2/3)\,.\label{Eq. 8}\end{equation}
 one can achieve via the SSB the following vacuum expectation values
(VEV) in the unitary gauge:

\begin{equation}
\left(\begin{array}{c}
\eta_{1}\left\langle \varphi\right\rangle +H_{\rho}\\
\\0\\
\\0\end{array}\right)\,,\quad\left(\begin{array}{c}
0\\
\\\eta_{2}\left\langle \varphi\right\rangle +H_{\chi}\\
\\0\end{array}\right)\,,\quad\left(\begin{array}{c}
0\\
\\0\\
\\\eta_{3}\left\langle \varphi\right\rangle +H_{\phi}\end{array}\right)\,,\label{Eq. 9}\end{equation}
with the overall VEV

\begin{equation}
\left\langle \varphi\right\rangle =\frac{\sqrt{\mu_{1}^{2}\eta_{1}^{2}+\mu_{2}^{2}\eta_{2}^{2}+\mu_{3}^{2}\eta_{3}^{2}}}{\sqrt{2\left(\lambda_{1}\eta_{1}^{4}+\lambda_{2}\eta_{2}^{4}+\lambda_{3}\eta_{3}^{4}\right)+\lambda_{4}\eta_{1}^{2}\eta_{2}^{2}+\lambda_{5}\eta_{1}^{2}\eta_{3}^{2}+\lambda_{6}\eta_{2}^{2}\eta_{3}^{2}}}\label{Eq. 10}\end{equation}
resulting from the minimum condition applied to the potential

\begin{equation}
\begin{array}{ccl}
V & = & \mu_{1}^{2}\rho^{+}\rho-\mu_{2}^{2}\chi^{+}\chi-\mu_{3}^{2}\phi^{+}\phi\\
\\ &  & +\lambda_{1}\left(\rho^{+}\rho\right)^{2}+\lambda_{2}\left(\chi^{+}\chi\right)^{2}+\lambda_{3}\left(\phi^{+}\phi\right)^{2}\\
\\ &  & +\lambda_{4}\left(\rho^{+}\rho\right)\left(\chi^{+}\chi\right)+\lambda_{5}\left(\rho^{+}\rho\right)\left(\phi^{+}\phi\right)+\lambda_{6}\left(\phi^{+}\phi\right)\left(\chi^{+}\chi\right)\\
\\ &  & +\lambda_{7}\left(\rho^{+}\chi\right)\left(\chi^{+}\rho\right)+\lambda_{8}\left(\rho^{+}\phi\right)\left(\phi^{+}\rho\right)+\lambda_{9}\left(\phi^{+}\chi\right)\left(\chi^{+}\phi\right).\end{array}\label{Eq. 11}\end{equation}

\section{Quasi-inverse seesaw mechanism}

With the above ingredients one can construct the Yukawa Ld allowed
by the gauge symmetry in the 3-3-1 model with right-handed neutrinos.
It simply is:

\begin{equation}
\begin{array}{ll}
\mathcal{-L}_{Y} & =h_{\phi}\overline{l}\phi e_{R}+h_{\rho}\overline{l}\rho N_{R}+h_{\chi}\overline{l}\chi N_{R}+\frac{1}{2}h_{\varphi}\overline{N_{R}^{c}}\eta_{0}\varphi N_{R}\\
\\ & +\frac{1}{2}h\varepsilon^{ijk}\left(\overline{l}\right)_{i}\left(l^{c}\right)_{j}\phi_{k}+h.c.\end{array}\label{Eq. 12}\end{equation}
where $h$s are $3\times3$ complex Yukawa matrices, the lower index
indicating the particular Higgs each one connects with.

It leads straightforwardly to the following mass terms:

\begin{equation}
\begin{array}{ll}
\mathcal{-L}_{mass} & =h_{\phi}\overline{e_{L}}e_{R}\left\langle \phi\right\rangle +h_{\rho}\overline{l}N_{R}\left\langle \rho\right\rangle +h_{\chi}\overline{l}N_{R}\left\langle \chi\right\rangle +\frac{1}{2}h_{\varphi}\overline{N_{R}}N_{R}^{c}\eta_{0}\left\langle \varphi\right\rangle \\
\\ & +\frac{1}{2}\left(h-h^{T}\right)\overline{\nu_{L}}\nu_{R}\left\langle \phi\right\rangle +h.c.\end{array}\label{Eq. 13}\end{equation}

The Yukawa terms allow one to construct the quasi-inverse seesaw mechanism
by displaying them into the following $9\times9$ complex matrix:

\begin{equation}
M=\left(\begin{array}{ccccc}
0 &  & \frac{1}{2}\left(h-h^{T}\right)\sqrt{\frac{1}{2\cos^{2}\theta_{W}}} &  & h_{\chi}\sqrt{\frac{a}{2}\left(1-\tan^{2}\theta_{W}\right)}\\
\\\frac{1}{2}\left(h^{T}-h\right)\sqrt{\frac{1}{2\cos^{2}\theta_{W}}} &  & 0 &  & h_{\rho}\sqrt{1-a}\\
\\h_{\chi}^{T}\sqrt{\frac{a}{2}\left(1-\tan^{2}\theta_{W}\right)} &  & h_{\rho}^{T}\sqrt{1-a} &  & h_{\varphi}\eta_{0}\end{array}\right)\left\langle \varphi\right\rangle \label{Eq. 14}\end{equation}

Due to the non-zero $h_{\chi}$ this matrix is slightly different
from the traditional inverse seesaw mechanism \cite{key-38} - \cite{key-40},
but its resulting effects - we prove in the following - are phenomenologically
plausible. However, this kind of seesaw matrix appears in the literature,
see for instance Refs. \cite{key-41,key-42}. This $9\times9$ complex
matrix can be displayed as:

\begin{equation}
M=\left(\begin{array}{ccc}
0 &  & m_{D}\\
\\m_{D}^{T} &  & M_{N}\end{array}\right)\label{Eq. 15}\end{equation}

with $m_{D}=\left(\begin{array}{ccc}
\frac{1}{2}\left(h-h^{T}\right)\sqrt{\frac{1}{2\cos^{2}\theta_{W}}} &  & h_{\chi}\sqrt{\frac{a}{2}\left(1-\tan^{2}\theta_{W}\right)}\end{array}\right)$ a $3\times6$ complex matrix and $M_{N}=\left(\begin{array}{cc}
0 & h_{\rho}\sqrt{1-a}\\
h_{\rho}\sqrt{1-a} & h_{\varphi}\eta_{0}\end{array}\right)$ a $6\times6$ complex matrix acting in the seesaw formula.

By diagonalizing the above matrix one gets the physical neutrino matrices
as: $M\left(\nu_{L}\right)\simeq-m_{D}\left(M_{N}^{-1}\right)m_{D}^{T}$
and $M\left(\nu_{R},N_{R}\right)\simeq M_{N}$which yield:

\begin{equation}
\begin{array}{cl}
M\left(\nu_{L}\right) & \simeq\frac{a\eta_{0}\left\langle \varphi\right\rangle }{8\left(1-a\right)\cos^{2}\theta_{W}}\left(h-h^{T}\right)\left(h_{\rho}^{-1}\right)^{T}\left(h_{\varphi}\right)\left(h_{\rho}^{-1}\right)\left(h^{T}-h\right)\\
\\ & -\frac{a\eta_{0}\sqrt{\left(1-\tan^{2}\theta_{W}\right)}\left\langle \varphi\right\rangle }{4\sqrt{\left(1-a\right)\cos^{2}\theta_{W}}}\left[\left(h_{\chi}\right)\left(h_{\rho}^{-1}\right)\left(h^{T}-h\right)+\left(h-h^{T}\right)\left(h_{\rho}^{-1}\right)^{T}\left(h_{\chi}\right)^{T}\right]\end{array}\label{Eq. 16}\end{equation}

\begin{equation}
\left(\begin{array}{cc}
M\left(\nu_{R}\right) & 0\\
\\0 & M\left(N_{R}\right)\end{array}\right)=\left(\begin{array}{cc}
h_{\rho}\sqrt{\left(1-a\right)}+\frac{1}{2}h_{\varphi}\eta_{0} & 0\\
\\0 & -h_{\rho}\sqrt{\left(1-a\right)}+\frac{1}{2}h_{\varphi}\eta_{0}\end{array}\right)\label{Eq. 17}\end{equation}

One can enforce here the realistic condition 

\begin{equation}
\left[\left(h_{\chi}\right)\left(h_{\rho}^{-1}\right)\left(h^{T}-h\right)\right]^{T}=-\left(h_{\chi}\right)\left(h_{\rho}^{-1}\right)\left(h^{T}-h\right)\label{Eq. 18}\end{equation}
in order to eliminate the troublesome terms in the left-handed neutrino
mass matrix. This condition can be naturally achieved if one takes
into consideration the plausible identity

\begin{equation}
h_{\chi}=h_{\rho}\label{Eq. 19}\end{equation}
meaning that the exotic right-handed neutrinos $N_{R}$ couples similarly
with $\nu_{L}$ and $\nu_{R}$respectively. Consequently, one gets
the left-handed neutrino mass matrix as the complex $3\times3$ matrix:

\begin{equation}
M\left(\nu_{L}\right)\simeq\frac{a\eta_{0}\left\langle \varphi\right\rangle }{8\left(1-a\right)\cos^{2}\theta_{W}}\left(h-h^{T}\right)\left(h_{\rho}^{-1}\right)^{T}\left(h_{\varphi}\right)\left(h_{\rho}^{-1}\right)\left(h^{T}-h\right)\label{Eq. 20}\end{equation}

It is evident that it is a pure Majorana mass matrix since $M\left(\nu_{L}\right){}^{T}=M\left(\nu_{L}\right)$
holds. Assuming that all the coupling matrices in the Yukawa sector
are of the same order of magnitude, say $\sim O(1)$, one can estimate
the order of magnitude of the individual masses in this matrix as

\begin{equation}
TrM\left(\nu_{L}\right)\simeq\frac{3a\eta_{0}\left\langle \varphi\right\rangle }{8\left(1-a\right)\cos^{2}\theta_{W}}\label{Eq. 21}\end{equation}

The right-handed neutrinos acquire some pseudo-Dirac masses, since
finally one remains with $M\left(\nu_{R}\right)^{T}\neq M\left(\nu_{R}\right)$
and $M\left(N_{R}\right)^{T}\neq M\left(N_{R}\right)$ and $h_{\rho}$
dictates their character.

\section{Tuning the parameters}

Now one can tune the parameters in this particular model in order
to get phenomenologically viable predictions. Obviously, both $a$,
$\eta_{0}\in(0,1)$. Since $\eta_{0}$is the parameter responsible
with the lepton number violation, one can keep it very small, say
$\eta_{0}\sim10^{-8}-10^{-9}$ in order to safely consider that the
global $U(1)_{leptonic}$ symmetry is very softly (quite negligible)
violated by the Majorana coupling it introduces. 

When comparing the boson mass spectrum in this model - obtained both
by using the general Cot\u{a}escu method \cite{key-26} and the SM
calculations \cite{key-1} - one gets a scales connection:

\begin{equation}
\sqrt{\left(1-\eta_{0}^{2}\right)a}=\frac{\left\langle \varphi\right\rangle _{SM}}{\left\langle \varphi\right\rangle }\label{Eq. 22}\end{equation}

It becomes obviously that $\eta_{0}$ has no part to play in the breaking
scales splitting. The later is determined quite exclusively by $a$.
If one takes $\left\langle \varphi\right\rangle _{SM}\simeq246$GeV
and $\left\langle \varphi\right\rangle \simeq1$TeV then $a\simeq0.06$.

With these plausible settings the individual neutrino masses come
out in the subsequent hierarchy:

\begin{equation}
\sum m\left(\nu_{L}\right)\simeq1eV\label{Eq. 23}\end{equation}

\begin{equation}
\sum m\left(\nu_{R}\right)\simeq\sum m\left(N_{R}\right)\simeq970GeV\label{Eq. 24}\end{equation}

Furthermore, one can enforce some extra flavor symmetries in the lepton
sector in order to dynamically get the appropriate PMNS mixing matrix.
Some discrete groups, such as $A_{4}$\cite{key-46,key-47}, $S_{4}$\cite{key-48}
or $S_{3}$\cite{key-49,key-50} can be employed in 3-3-1 models with
no exotic electric charges, in order to accomplish this task, but
this exceeds the scope of this letter.

\section{Conclusions}

We have discussed here the possible realization of a quasi-inverse
seesaw mechanism in the 3-3-1 class of gauge models with ''lepton
number violating'' exotic Majorana neutrinos added. The Cot\u{a}escu
general method of treating gauge models with high-symmetries is involved
and it successfully provides us not only with the one-parameter mass
spectrum but also with the lepton number violating terms needed for
a plausible inverse seesaw mechanism due to the possivility of coupling
the $\varphi$ to exotic Majorana neutrinos. To the extent of our
knowledge, in low energy models one finds no such terms to give masses
to exotic neutrinos, so that some extra assumptions (usually from
GUT theories) are invoked. These two characteristics single out our
approach from other recent similar attempts \cite{key-34,key-35}.
The details of the mixing in the neutrino sector are closely related
to the entries in $h$, $h_{\rho}$ and $h_{\varphi}$ but this lies
beyond the scope of this letter and will be presented elsewhere. The
framework of this kind of 3-3-1 models is a very promising one. It
recovers all the results of the SM and in addition exhibits a lot
of assets: it requires precisely 3 fermion generations, its algebraic
structure dictates the observed charge quantization, it can predict
a testable Higgs phenomenology and, as we presented here, is suitable
for neutrino phenomenology.

\section*{Acknowledgment}

The author would like to thank Prof. I. I. Cot\u{a}escu for fruitful
discussions on topics included in this letter.


\begin{thebibliography}{10}
\bibitem{key-1}J. F. Donoghue, E. Golowich and B. R. Holstein, \emph{Dynamics of
the Standard Model}, (second edition - Cambridge University Press,
2014). 
\bibitem{key-2}C. Quigg, G\emph{auge Theories of the Strong, Weak and Electromagnetic
Interactions} (second edition - Princeton University Press, 2013).
\bibitem{key-3}Ta-Pei Cheng and Ling-Fong Li, \emph{Gauge theory of elementary particle
physics} (Oxford University Press, 1988).
\bibitem{key-4}J. Beringer \emph{et al}. (PDG), \emph{Phys. Rev. D} \textbf{86},
010001 (2012). 
\bibitem{key-5}A. Strumia and F. Vissani, arXiv: hep-ph/0606054 v3. P
\bibitem{key-6}P. H. Frampton, \emph{Phys. Rev. Lett.} \textbf{69}, 2889 (1992). 
\bibitem{key-7}F. Pisano and V. Pleitez, \emph{Phys. Rev.} \emph{D} \textbf{46},
410 (1992). 
\bibitem{key-8}R. Foot, H. N. Long and T. A. Tran, \emph{Phys. Rev.} \emph{D} \textbf{50},
R34 (1994). 
\bibitem{key-9}H. N. Long, \emph{Phys. Rev.} \emph{D} \textbf{53}, 437 (1996). 
\bibitem{key-10}W. A. Ponce, J. B. Florez and L. A. Sanchez, \emph{Int. J. Mod. Phys.
A} \textbf{17}, 643 (2002). 
\bibitem{key-11}L. A. Sanchez, W. A. Ponce and R. Martinez, \emph{Phys. Rev.} \emph{D}
\textbf{64}, 075013 (2001). 
\bibitem{key-12}R. A. Diaz, R. Martinez and F. Ochoa, \emph{Phys. Rev.} \emph{D} \textbf{72},
035018 (2005).
\bibitem{key-13}G. Tavares-Velasco and J. J. Toscano, \emph{Phys. Rev.} \emph{D} \textbf{70},
053006 (2004).
\bibitem{key-14}A. Doff, C. A. da S. Pires and P. S. Rodrigues da Silva, \emph{Phys.
Rev.} \emph{D} \textbf{74}, 015014 (2006). 
\bibitem{key-15}A. Carcamo, R. Martinez and F. Ochoa, \emph{Phys. Rev.} \emph{D} \textbf{73},
035007 (2006). 
\bibitem{key-16}F. Ramirez-Zavaleta, G. Tavares-Velasco and J. J. Toscano, \emph{Phys.
Rev.} \emph{D} \textbf{75}, 075008 (2007). 
\bibitem{key-17}E. Ramirez-Barreto, Y. A. Coutinho and J. Sa Borges, \emph{Eur. Phys.
J.} \emph{C} \textbf{50}, 909 (2007).
\bibitem{key-18}A. Palcu, \emph{Mod. Phys. Lett.} \emph{A} \textbf{23}, 387 (2008). 
\bibitem{key-19}J. M. Cabarcas, D. Gomez Dumm and R. Martinez, \emph{Eur. Phys. J.}
\emph{C} \textbf{58}, 569 (2008). 
\bibitem{key-20}R. Martinez and F. Ochoa, \emph{Phys. Rev. D} \textbf{77}, 065012
(2008).
\bibitem{key-21}R. H. Benavides, Y. Giraldo and W. A. Ponce, \emph{Phys. Rev. D} \textbf{80},
113009 (2009). 
\bibitem{key-22}C. Alvarado, R. Martinez and F. Ochoa, \emph{Phys. Rev. D} \textbf{86},
025027 (2012). 
\bibitem{key-23}D. Cogollo et al., \emph{Eur. Phys. J.} \emph{C} \textbf{72}, 2029
(2012)., 
\bibitem{key-24}A. G. Dias, C. A. de S. Pires and P. S. Roridgues da Silva, \emph{Phys.
Lett. B} \textbf{628}, 85 (2005).
\bibitem{key-25}D. Chang and H. N. Long, \emph{Phys. Rev.} \emph{D} \textbf{73}, 053006
(2006).
\bibitem{key-26}A. Palcu, \emph{Mod. Phys. Lett.} \emph{A} \textbf{21}, 1203 (2006).
\bibitem{key-27}A. Palcu, \emph{Mod. Phys. Lett.} \emph{A} \textbf{21}, 2027 (2006). 
\bibitem{key-28}A. Palcu, \emph{Mod. Phys. Lett.} \emph{A} \textbf{21}, 2591 (2006).
\bibitem{key-29}A. Palcu, \emph{Mod. Phys. Lett.} \emph{A} \textbf{22}, 939 (2007).
\bibitem{key-30}P. V. Dong and H. N. Long, \emph{Phys. Rev.} \emph{D} \textbf{77},
057302 (2008).
\bibitem{key-31}D. Cogollo, H. Diniz, C. A. de S. Pires and P. S. Rodrigues da Silva,
\emph{Eur. Phys. J. C} \textbf{58}, 455 (2008).
\bibitem{key-32}D. Cogollo, H. Diniz and C. A. de S. Pires, \emph{Phys. Lett B} \textbf{677},
338 (2009).
\bibitem{key-33}D. Cogollo, H. Diniz and C. A. de S. Pires, \emph{Phys. Lett B} \textbf{687},
400 (2010).
\bibitem{key-34}A. G. Dias C. A. de S. Pires, P. S. Rodrigues da Silva and A. Sampieri,
\emph{Phys. Rev. D} \textbf{86}, 035007 (2012).
\bibitem{key-35}E. Catano M., R. Martinez and F. Ochoa, \emph{Phys. Rev. D} \textbf{86},
073015 (2012).
\bibitem{key-36}I. I. Cot\u{a}escu, \emph{Int. J. Mod. Phys. Rev.} \emph{A} \textbf{12},
1483(1997).
\bibitem{key-37}A. Palcu, \emph{Prog.Theor. Exp Phys}. \textbf{2013}, 0903B03 (2013).
\bibitem{key-38}R. Mohapatra, \emph{Phys. Rev. Lett} \textbf{56}, 561 (1986).
\bibitem{key-39}R. Mohapatra and J. Valle, \emph{Phys. Rev. D} \textbf{34}, 1642 (1986).
\bibitem{key-40}P.-H. Gu and U. Sarkar, \emph{Phys. Lett B} \textbf{694}, 226 (2010).
\bibitem{key-41}A. Ibarra, E. Molinaro and S. Petcov, \emph{JHEP} \textbf{09}, 108
(2010).
\bibitem{key-42}S. Barr, \emph{Phys. Rev. Lett} \textbf{92}, 101 (2004).
\bibitem{key-43}P.S. Bhupal Dev and R. N. Mohapatra, \emph{Phys. Rev. D} \textbf{81},
013001 (2010).
\bibitem{key-44}A. Das and N. Okada, \emph{Phys. Rev. D} \textbf{88}, 113001 (2013)
\bibitem{key-45}A. Das, P.S. Bhupal Dev and N. Okada, \emph{Phys. Lett. B} \textbf{735},
364 (2014). 
\bibitem{key-46}Furong Yin, \emph{Phys. Rev. D} \textbf{75}, 073010 (2007).
\bibitem{key-47}P. V. Dong et al.,\emph{Phys. Rev. D} \textbf{81}, 053004 (2010).
\bibitem{key-48}P. V. Dong et al.\emph{, Eur. Phys. J. C} \textbf{71}, 1544 (2011).
\bibitem{key-49}P. V. Dong et al., \emph{Phys. Rev. D} \textbf{85}, 053001 (2012).
\bibitem{key-50}A. E. Carcamo Hernandez, E. Catano Mur and R. Martinez, arXiv: 1407.5217
{[}hep-ph{]}.\end{thebibliography}
\end{document}